\documentstyle[12pt,epsf]{article}

\renewcommand{\pmod}[1]{\ ({\bmod}\ #1)}

\newcommand{\ket}[1]{|#1\rangle}
\newcommand{\Pf}{\mathop{\rm Pf}\nolimits}
\newcommand{\sgn}{\mathop{\rm sgn}\nolimits}
\newcommand{\calM}{{\cal M}}

\newcommand{\lesssim}{\mathrel{\mathop{\mathchoice%
{\normalsize\raisebox{0.3ex}[1.1ex][-0.6ex]{$<$}}%
{\normalsize\raisebox{0.3ex}[1.1ex][-0.6ex]{$<$}}%
{\scriptsize\raisebox{0.3ex}[1.0ex][-0.8ex]{$<$}}%
{\scriptsize\raisebox{0.3ex}[1.0ex][-0.8ex]{$<$}}}%
\limits_\sim}}

\newenvironment{subfig}[1]%
{\def\subfigLabel{#1}\begin{tabular}[b]{@{}c@{}}}%
{\\[10pt]\subfigLabel\end{tabular}}

\newsavebox{\TempBox}
\newlength{\TempLength}

\title{Unpaired Majorana fermions in quantum wires}
\author{Alexei~Yu.~Kitaev\thanks{On leave from  L.~D.~Landau Institute
for Theoretical Physics}\\
{\it Microsoft Research}\\
{\it Microsoft, \#113/2032, One Microsoft Way,}\\
{\it Redmond, WA 98052, U.S.A.}\\
{\it kitaev@microsoft.com}
}
\date{27 October 2000}

\begin{document}

\maketitle

\begin{abstract}
Certain one-dimensional Fermi systems have an energy gap in the bulk spectrum
while boundary states are described by one Majorana operator per boundary
point. A finite system of length $L$ possesses two ground states with an
energy difference proportional to $\exp(-L/l_0)$ and different fermionic
parities. Such systems can be used as qubits since they are intrinsically
immune to decoherence. The property of a system to have boundary Majorana
fermions is expressed as a condition on the bulk electron spectrum. The
condition is satisfied in the presence of an arbitrary small energy gap
induced by proximity of a 3-dimensional $p$-wave superconductor, provided that
the normal spectrum has an odd number of Fermi points in each half of the
Brillouin zone (each spin component counts separately).
\end{abstract}

\section*{Introduction}

Implementing a full scale quantum computer is a major challenge to modern
physics and engineering. Theoretically, this goal should be achievable due to
the possibility of fault-tolerant quantum computation \cite{Shor}. Unlimited
quantum computation is possible if errors in the implementation of each gate
are below certain threshold~\cite{AB1,AB2,KLZ,Kit1}. Unfortunately, for
conventional fault-tolerance schemes the threshold appears to be about
$10^{-4}$, which is beyond the reach of current technologies. It has been also
suggested that fault-tolerance can be achieved at the physical level (instead
of using quantum error-correcting codes). The first proposal of these
kind~\cite{Kit2} was based on non-Abelian anyons in two-dimensional systems.
The relation between quantum computation, anyons and topological quantum field
theories was independently discussed in~\cite{Freedman}. A mathematical result
about universal quantum computation with certain type of anyons has been
obtained recently~\cite{FLW}, but, generally, this approach is still
undeveloped. In these paper we describe another (theoretically, much simpler)
way to construct decoherence-protected degrees of freedom in one-dimensional
systems (``quantum wires''). Although it does not automatically provide
fault-tolerance for {\em quantum gates}, it should allow, when implemented, to
build a reliable {\em quantum memory}.

The reason why quantum states are so fragile is that they are sensitive to
errors of two kinds. A classical error, represented by an operator
$\sigma^x_j$, flips the $j$-th qubit changing $|0\rangle$ to $|1\rangle$ and
vice versa. A phase error $\sigma^z_j$ changes the sign of all states with the
$j$-th qubit equal to $1$ (i.\,e.\ $j$-th spin down, if the qubits are spins)
relative to the states with the $j$-th qubit equal to $0$. It is generally
easy to get rid of one type of errors, but not both. However, the following
method of eliminating the classical errors is worth considering.  Let each
qubit be a site that can be either empty or occupied by an electron (with spin
up, say, the other spin direction being forbidden). Let us denote the empty
and the occupied states by $|0\rangle$ and $|1\rangle$, respectively. (Such
sites are not exactly qubits because electrons are fermions, but they can be
also used for quantum computation~\cite{BK}). Now single classical errors
become impossible because the electric charge is conserved. Even in
superconducting systems, the fermionic parity (i.\,e.\ the electric
charge$\pmod{2}$) is conserved. Two classical errors can still happen at two
sites simultaneously, but this would require that an electron jumps from one
site to the other. Such jumps can be avoided by placing the ``fermionic
sites'' far apart from each other, provided the medium between them has an
energy gap in the excitation spectrum.

Obviously, this method does not protect from phase errors which are now
described by the operators $a_j^\dagger a_j$. To the contrary, different
electron configurations will have different energies and thus will pick up
different phases over time. Even without actual inelastic processes, this will
produce the same effect as decoherence. However, a simple mathematical
observation suggests that the situation could be improved. Each fermionic site
is described by a pair of annihilation and creation operators $a_j$,
$a_j^\dagger$. One can formally define {\em Majorana operators}
\begin{equation}\label{Majorana0}
c_{2j-1}=a_j+a_j^\dagger, \qquad c_{2j}=\frac{a_j-a_j^\dagger}{i}
\qquad\quad (j=1,\ldots,N)
\end{equation}
which satisfy the relations
\begin{equation}\label{Majorana_relations}
c_m^\dagger=c_m, \qquad c_lc_m+c_mc_l=2\delta_{lm}
\qquad\quad (l,m=1,\ldots,2N).
\end{equation}
If the operators $c_{2j-1}$ and $c_{2j}$ belonged to
different sites then the phase error $a_j^\dagger a_j=
\frac{1}{2}(1+ic_{2j-1}c_{2j})$ would be unlikely to occur. Indeed, it would
require interaction between the two ``Majorana sites'' which could be possibly
avoided. Note that a single Majorana operator $c_{2j-1}$ or $c_{j}$ can not
appear as a term in any reasonable Hamiltonian because it does not preserve
the fermionic parity. Thus an isolated Majorana site (usually called a {\em
Majorana fermion}) is immune to any kind of error!

Unfortunately, Majorana fermions are not readily available in solid state
systems. The goal of this paper is to construct Hamiltonians which would give
rise to Majorana fermions as effective low-energy degrees of freedom.
Surprisingly, this can be done even with non-interacting electrons. (Some
interaction is actually needed to create superconductivity, but it can be
effectively described by terms like $\Delta a_ja_{k}$). The general idea is
quite simple. An arbitrary quadratic Hamiltonian can be written in the form
\begin{equation}\label{quadratic_Hamiltonian}
H= \frac{i}{4}\sum_{l,m} A_{lm}c_lc_m \qquad\quad
(A_{lm}^*=A_{lm}=-A_{ml}).
\end{equation}
Its ground state can be described as ``pairing'' of Majorana operators: normal
mode creation and annihilation operators $\tilde a_m^\dagger,\tilde a_m$,
which are certain linear combinations of $c_l$, come in pairs. (In this
sense, an insulator and a superconductor represent different types of
pairing). In some cases, most Majorana operators are paired up with an energy
gap while few ones (localized at the boundary or defects) remain ``free''. For
example, unpaired Majorana fermions exist on vortices in chiral 2-dimensional
$p$-wave superconductors~\cite{ReadGreen,Ivanov}. We will show that Majorana
fermions can also occur at the ends of quantum wires.

\section{A toy model and the qualitative picture} \label{sec_model}

\begin{figure}[ht]
\hbox to \textwidth{\hfill

\begin{picture}(110,35)
\put(0,0){\line(1,0){100}}
\put(0,15){\line(1,0){100}}
\put(10,35){\line(1,0){100}}
\put(0,0){\line(0,1){15}}
\put(100,0){\line(0,1){15}}
\put(110,20){\line(0,1){15}}
\put(0,15){\line(1,2){10}}
\put(100,15){\line(1,2){10}}
\put(100,0){\line(1,2){10}}
\put(15,24.7){\thicklines\line(1,0){80}}
\put(15,25.3){\thicklines\line(1,0){80}}
\put(8,22){\footnotesize$b'$}
\put(96,22){\footnotesize$b''$}
\put(0,4){\hbox to 100\unitlength {\footnotesize\hfil$\theta$\hfil}}
\end{picture}

\hfill}
\caption{A piece of ``quantum wire'' on the surface of 3-dimensional
superconductor.}
\label{fig_layout}
\end{figure}
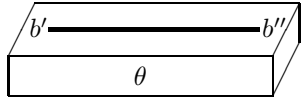

We are going to describe a simple but rather unrealistic model which exhibits
unpaired Majorana fermions. It attempts to catch two important properties
which seem necessary for the phenomenon to occur. Firstly, the $U(1)$ symmetry
$a_j\mapsto e^{i\phi}a_j$, corresponding to the electric charge conservation,
must be broken down to a ${\bf Z}_2$ symmetry, $a_j\mapsto -a_j$. Indeed, if a
single Majorana operator can be localized, symmetry transformation should not
mix it with other operators. So we should consider superconductive systems.
The particular mechanism of superconductivity is not important; we may just
think that our quantum wire lies on the surface of 3-dimensional
superconductor (see fig.~\ref{fig_layout}). The second property is less
obvious and will be fully explained in Sec.~\ref{sec_condition}. Roughly
speaking, the electron spectrum must strongly depend on the spin. Here we will
simply assume that only one spin component (say, $\uparrow$) is
present.\,\footnote{\label{remark_triplet} It appears that only a triplet
($p$-wave) superconductivity in the 3-dimensional substrate can effectively
induce the desired pairing between electrons with the same spin direction ---
at least, this is true in the absence of spin-orbit interaction.}

Consider a chain consisting of $L\gg 1$ sites. Each site can be either
empty or occupied by an electron (with a fixed spin direction).  The
Hamiltonian is
\begin{equation}\label{the_model}
H_1\,=\, \sum_j \Bigl( -w(a_j^\dagger a_{j+1}+a_{j+1}^\dagger a_j)
-\mu(a_j^\dagger a_j-{\textstyle\frac{1}{2}})
+\Delta a_ja_{j+1}+\Delta^* a_{j+1}^\dagger a_j^\dagger \Bigr).
\end{equation}
Here $w$ is a hopping amplitude, $\mu$ a chemical potential, and
$\Delta=|\Delta|e^{i\theta}$ the induced superconducting gap. It is convenient
to hide the dependence on the phase parameter $\theta$ into the definition of
Majorana operators:
\begin{equation}\label{Majorana}
c_{2j-1}=e^{i\frac{\theta}{2}}a_j+e^{-i\frac{\theta}{2}}a_j^\dagger, \quad\
c_{2j}=-ie^{i\frac{\theta}{2}}a_j+ie^{-i\frac{\theta}{2}}a_j^\dagger
\qquad (j=1,\ldots,L).
\end{equation}
In terms of this operators, the Hamiltonian becomes
\begin{equation}\label{the_model_Majorana}
H_1\,=\, \frac{i}{2} \sum_j \Bigl( -\mu c_{2j-1}c_{2j}
+(w+|\Delta|)c_{2j}c_{2j+1}+(-w+|\Delta|)c_{2j-1}c_{2j+2} \Bigr).
\end{equation}

\begin{figure}[ht]

\sbox{\TempBox}{%
\begin{picture}(0,0)
\put(0,0){\oval(32,16)}
\put(-10,0){\circle*{4}}
\put(10,0){\circle*{4}}
\end{picture}}

\newcommand\fsite[2]{%
\begin{picture}(0,0)
\put(0,0){\usebox{\TempBox}}
\put(-16,12){\footnotesize #1}
\put(10,12){\footnotesize #2}
\end{picture}}

\hbox to \textwidth {

\begin{subfig}{a)}
\begin{picture}(170,30)(0,-8)
\put(18,0){\fsite{$c_1$}{$c_2$}}
\put(8,0){\thicklines\line(1,0){20}}
\put(75,0){\fsite{$c_3$}{$c_4$}}
\put(65,0){\thicklines\line(1,0){20}}
\put(104,0){\ldots}
\put(146,0){\fsite{$c_{2L-1}$}{$c_{2L}$}}
\put(136,0){\thicklines\line(1,0){20}}
\end{picture}
\end{subfig}

\hfill

\begin{subfig}{b)}
\begin{picture}(170,50)(0,-8)
\put(18,0){\fsite{$c_1$}{$c_2$}}
\put(28,0){\thicklines\line(1,0){37}}
\put(75,0){\fsite{$c_3$}{$c_4$}}
\put(85,0){\thicklines\line(1,0){18}}
\put(104,0){\ldots}
\put(146,0){\fsite{$c_{2L-1}$}{$c_{2L}$}}
\put(136,0){\thicklines\line(-1,0){18}}
\end{picture}
\end{subfig}

}

\caption{Two types of pairing.}\label{fig_pairing}
\end{figure}
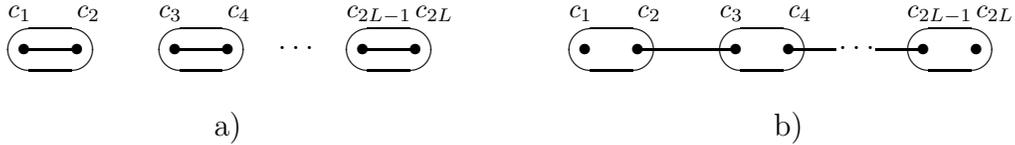

Let us start with two special cases.
\begin{itemize}
\item[a)] The trivial case: $|\Delta|=w=0$,\, $\mu<0$. Then $H_1=-\mu\sum_j
(a_j^\dagger a_j-\frac{1}{2})=\frac{i}{2}(-\mu)\sum_j c_{2j-1}c_{2j}$. The
Majorana operators $c_{2j-1},c_{2j}$ from the same site $j$ are paired
together to form a ground state with the occupation number $0$.
\item[b)] $|\Delta|=w>0$,\, $\mu=0$. In this case
\begin{equation}
H_1=iw\sum_j c_{2j}c_{2j+1}.
\end{equation}
Now the Majorana operators $c_{2j},c_{2j+1}$ {\em from different sites} are
paired together (see fig.~\ref{fig_pairing}). One can define new annihilation
and creation operators $\tilde a_j=\frac{1}{2}(c_{2j}+ic_{2j+1}),\, \tilde
a_j^\dagger=\frac{1}{2}(c_{2j}-ic_{2j+1})$ which span the sites $j$ and
$j+1$. The Hamiltonian becomes $2w\sum_{j=1}^{L-1}(\tilde a_j^\dagger\tilde
a_j-\frac{1}{2})$. Ground states satisfy the condition $\tilde
a_j\ket{\psi}=0$ for $j=1,\ldots,L-1$. There are two orthogonal states
$\ket{\psi_0}$ and $\ket{\psi_1}$ with this property. Indeed, the Majorana
operators $b'=c_1$ and $b''=c_{2L}$ remain unpaired (i.\,e.\ do not enter the
Hamiltonian), so we can write
\begin{equation}\label{ground_states}
-ib'b''\ket{\psi_0}=\ket{\psi_0},\qquad -ib'b''\ket{\psi_1}=-\ket{\psi_1}.
\end{equation}
\end{itemize}
\vspace{-7mm}

\noindent
Note that the state $\ket{\psi_0}$ has an even fermionic parity (i.\,e.\ it is
a superposition of states with even number of electrons) while $\ket{\psi_1}$
has an odd parity. The parity is measured by the operator
\begin{equation}\label{parity_operator}
P= \prod_{j}(-ic_{2j-1}c_{2j}).
\end{equation}

These two cases represent two phases, or universality classes which exist in
the model. A subtle point is that both phases have the same bulk
properties. In fact, one phase can be transformed to the other (and vice
versa) by mere permutation of Majorana operators,
\begin{equation}\label{shift}
c_m\mapsto c_{m+1}.
\end{equation}
Such a local transformation (operator algebra automorphism) is usually
considered as ``equivalence'' in the study of lattice models.\,\footnote{
Nonlocal transformations can change the physical properties of the model even
more dramatically. The Jordan-Wigner transformation
$c_{2j-1}\mapsto\sigma^x_j\prod_{k=1}^{j-1}\sigma^z_k$,\,\,
$c_{2j}\mapsto\sigma^y_j\prod_{k=1}^{j-1}\sigma^z_k$\, takes our model to a
spin chain with $xx$ and $yy$ interactions and a $z$-directed magnetic field.
Unlike~(\ref{shift}), the Jordan-Wigner transformation is well defined at the
ends of the chain. However, this mathematical procedure falls apart in the
physical context, as far as perturbations are involved. Indeed, the phase (b)
has now an order parameter $\langle\sigma^x\rangle\not=0$. External fields
will interact with the order parameter breaking the phase coherence between
$\ket{\psi_0}$ and $\ket{\psi_1}$.} Yet the boundary properties of the two
phases are clearly different: only the phase (b) has unpaired Majorana
fermions at the ends of the chain. This is due to the fact that the operators
$c_{2j-1},c_{2j}$ belong to one physical site while $c_{2j},c_{2j+1}$ do
not. We may put it this way: one can not cut a physical site into two halves;
if one could, both types of boundary states would be possible in both
phases.

Also note that the transformation~(\ref{shift}) can not be performed in a
continuous fashion, starting from the identity transformation. From the
mathematical perspective, it means that one should have different definitions
for ``weak'' and ``strong'' equivalence of lattice models . We
will not touch such abstract matters here.

Now we want to study the model at arbitrary values of $w$, $\mu$ and $\Delta$.
Let us begin with some generalities. Let $N$ be the total number
of fermionic sites in the system, for now $N=L$. The
Hamiltonian~(\ref{the_model_Majorana}) has the general
form~(\ref{quadratic_Hamiltonian}). Hence it can be reduced to a canonical
form
\begin{equation}\label{canonical_form}
H_{\rm canonical} \,=\,\frac{i}{2}\sum_{m=1}^{N}\epsilon_m b_m'b_m'' \,=\,
\sum_{m=1}^{N}\epsilon(\tilde a_m^\dagger\tilde a_m-{\textstyle\frac{1}{2}})
\qquad\quad (\epsilon_m\ge 0).
\end{equation}
Here $b_m',b_m''$ are real linear combinations of $c_{2j-1},c_{2j}$ with the
same commutation relations whereas $\tilde a_m=\frac{1}{2}(b_m'+ib_m'')$,\,
$\tilde a_m^\dagger=\frac{1}{2}(b_m'-ib_m'')$. More specifically,
\begin{equation}\label{reduction_to_canonical}
\left(\begin{array}{c} b_1'\\ b_1''\\ \vdots\\ b_N'\\ b_N''\end{array}\right)
= W \left(\begin{array}{c}
  c_1\\ c_2\\ \vdots\\ c_{2N-1}\\ c_{2N}\end{array}\right),
\qquad\quad
WAW^T = \left(\begin{array}{ccccc}
  0 & \epsilon_1 &&&\\ -\epsilon_1 & 0 &&&\\ &&\ddots&&\\
  &&& 0 & \epsilon_N\\ &&& -\epsilon_N & 0 \end{array}\right),
\end{equation}
where $W$ is a $2N\times 2N$ real orthogonal matrix ($W^TW=WW^T=I$) whose rows
are eigenvectors of $A$. The numbers $\epsilon_m\ge 0$ are one-particle
excitation energies. However, it is more convenient to deal with a ``double
spectrum'' $\{\epsilon_m,-\epsilon_m\}$ since the matrix $A$ has eigenvalues
$\pm i\epsilon_m$.

The bulk spectrum (energy vs. momentum) is given by
\begin{equation}
\epsilon(q)=\pm\sqrt{(2w\cos q+\mu)^2+4|\Delta|^2\sin^2q}, \qquad\quad
-\pi\le q\le \pi.
\end{equation}
We may conjecture that the phases (a) and (b) extend to connected domains in
the parameter space where the spectrum has a gap. The signs of $\mu$ and $w$
seem not to be important, so we actually expect that the phase (a) occurs at
$2|w|<|\mu|$ while the phase (b) occupies the domain $2|w|>|\mu|$,
$\Delta\not=0$. (The phase boundary is given by the equation $2|w|=|\mu|$
while $\Delta=0$,\, $2|w|>|\mu|$ is a line of normal metal phase inside the
domain (b)).

To verify the conjecture, we need to find boundary modes. They correspond to
eigenvectors of $A$ localized near the ends of the chain. Due to the spectrum
symmetry $\epsilon\mapsto-\epsilon$, zero eigenvalues can occur in a general
position. If exist, such zero modes should have the form
\begin{equation}\label{zero_modes}
\begin{array}{rcl}
b'&=& \sum_j (\alpha'_{+}x_{+}^j+\alpha'_{-}x_{-}^j)\, c_{2j-1}
\medskip\\
b''&=& \sum_{j} (\alpha''_{+}x_{+}^{-j}+\alpha''_{-}x_{-}^{-j})\, c_{2j}
\end{array}
\qquad\quad
x_{\pm}=\frac{-\mu\pm\sqrt{\mu^2-4w^2+4|\Delta|^2}}{2(w+|\Delta|)}
\end{equation}
We will consider two cases corresponding to the expected existence domains of
the two phases.
\begin{itemize}
\item[a)] If $2|w|<|\mu|$, we have $|x_+|>1$, $|x_-|<1$ or $|x_+|<1$,
$|x_-|>1$. Therefore, only one of the coefficients $\alpha'_{+},\alpha'_{-}$
(or $\alpha''_{+},\alpha''_{-}$) can be non-zero, depending on whether the
mode is to be localized at the left or at the right end of the chain. This
makes it impossible to satisfy boundary conditions. So the supposed zero
modes~(\ref{zero_modes}) do not exist.
\item[b)] If $2w>|\mu|$,\, $\Delta\not=0$, we find that $|x_+|,|x_-|<1$. Hence
$b'$ is localized near $j=0$ whereas $b''$ is localized near $j=L$. There are
also boundary conditions $\alpha'_{+}+\alpha'_{-}=0$,\,
$\alpha''_{+}x_{+}^{-(L+1)}+\alpha''_{-}x_{-}^{-(L+1)}=0$, but they can be
satisfied too. The zero modes $b',b''$ are actually the same as the unpaired
Majorana fermions discussed before. If $-2w>|\mu|$,\, $\Delta\not=0$ then $b'$
and $b''$ change places. Thus the unpaired Majorana fermions exist in the
whole expected domain of the phase (b).
\end{itemize}

The above analysis is exact in the limit $L\to\infty$. If the chain length $L$
is finite, there is a weak interaction between $b'$ and $b''$.  (For
definiteness, we will always assume that $b'$ is at the left end of the chain
whereas $b''$ is at the right end). This interaction is described by an
effective Hamiltonian
\begin{equation}\label{effective_Hamiltonian}
H_{\rm eff}=\frac{i}{2}\,t\,b'b'',\qquad\quad t\propto e^{-L/l_0},
\end{equation}
where $l_0^{-1}$ is the smallest of $\Bigl|\ln|x_{+}|\Bigr|$ and
$\Bigl|\ln|x_{-}|\Bigr|$ (note that both logarithms have the same sign).  Thus
the energies of the ground states $\ket{\psi_0}$ and $\ket{\psi_1}$ (see
eq.~(\ref{ground_states})) differ by $t$. Note that it is not obvious any more
which state of the two is even and which is odd. In the case $-2w>|\mu|$, the
parity is proportional to $(-1)^L$. (This factor is the parity of the bulk
part of the chain).

The effective Hamiltonian~(\ref{effective_Hamiltonian}) still holds if we
include small electron-electron interaction (a four-fermion term)
into~(\ref{the_model}).  Indeed, the physical meaning of $t$ is an amplitude
for a fermionic quasiparticle to tunnel across the chain. In a long chain,
this amplitude vanishes as $e^{-L/l_0}$ if the bulk spectrum has a gap.

Finally, we will discuss a role of the phase parameter $\theta$\,
($\Delta=e^{i\theta}|\Delta|$). According to eq.~(\ref{Majorana}), the
Majorana operators $c_{2j-1},c_{2j}$ are multiplied by $-1$ when $\theta$
changes by $2\pi$. The physical parameter $\Delta$ is the same at $\theta$ and
$\theta+2\pi$, of course, but the ground states should undergo certain
transformation as $\theta$ changes to $\theta+2\pi$ adiabatically. Note that
the transformation $c_m\mapsto-c_m$ also occurs if one conjugates $c_m$ by the
parity operator $P$. Within the effective Hamiltonian approach, $P$ is the
same as $s(L)(-ib'b'')$\, ($s(L)=\pm 1$). Hence the adiabatic change of the
superconducting phase by $2\pi$ results in the unitary transformation
\begin{equation}\label{V}
V=s(L)(-ib'b'')\ :\qquad\
V\ket{\psi_0}=\ket{\psi_0},\quad\
V\ket{\psi_1}=-\ket{\psi_1}.
\end{equation}
This is equivalent to transfer of an electron between the ends of the
chain. Some physical consequences of this result will be mentioned in
Sec.~\ref{sec_realization}.

\section{A general condition for Majorana fermions}\label{sec_condition}

Let us consider a general translationally invariant one-dimensional
Hamiltonian with short-range interactions. It has been mentioned that the
necessary conditions for unpaired Majorana fermions are superconductivity and
a gap in the bulk excitation spectrum. The latter is equivalent to the
quasiparticle tunneling amplitude vanishing as $e^{-L/l_0}$. Besides that, it
is clear that there should be some parity condition. Indeed, Majorana fermions
at the ends of parallel weakly interacting chains may pair up and cancel each
other (i.\,e\ the ground state will be non-degenerate). So, provided the
energy gap, each one-dimensional Hamiltonian $H$ is characterized by a
``Majorana number'' $\calM=\calM(H)=\pm 1$: the existence of unpaired Majorana
fermions is indicated as $\calM=-1$. The Majorana number should satisfy
$\calM(H'\oplus H'')=\calM(H')\calM(H'')$, where $\oplus$ means taking two
non-interacting chains.

Remarkably, the Majorana number reveals itself even if the chain is closed
into a loop. This is handy as it eliminates the need to study boundary modes.
Let $H(L)$ be the Hamiltonian of a closed chain of length $L\gg l_0$. ($H$
itself is a template which is used to generate $H(L)$ for any $L$). We claim
that
\begin{equation}\label{loop_condition}
P(H(L_1+L_2))\,=\,\calM(H)\,P(H(L_1))\,P(H(L_2)),
\end{equation}
where $P(X)$ denotes the ground state parity of a Hamiltonian $X$
(assuming that the ground state is unique).

\begin{figure}[ht]

\sbox{\TempBox}{%
\begin{picture}(0,0)
\put(0,0){\circle{6}}
\put(-6,0){\thicklines\line(1,0){3}}
\put(6,0){\thicklines\line(-1,0){3}}
\end{picture}}

\newcommand{\chain}[4]{%
\TempLength=10\unitlength%
\TempLength=#1\TempLength%
\addtolength{\TempLength}{10\unitlength}%
\begin{picture}(0,0)
\multiput(6,0)(10,0){#1}{\usebox{\TempBox}}
\put(-3,8){\hbox to \TempLength {\footnotesize\hfil#2\hfil}}
\put(-3,-13){\hbox to \TempLength {\footnotesize#3\hfil#4}}
\end{picture}}

\hbox to \textwidth{\hfill

\begin{subfig}{a)}
\begin{picture}(150,36)(-3,-16)
\put(0,0){\chain{6}{$L_1$}{$b_1'$}{$b_1''$}}
\put(90,0){\chain{5}{$L_2$}{$b_2'$}{$b_2''$}}
\put(0,0){\line(0,1){20}}
\put(62,0){\line(0,1){20}}
\put(90,0){\line(0,1){20}}
\put(142,0){\line(0,1){20}}
\put(0,20){\line(1,0){62}}
\put(90,20){\line(1,0){52}}
\end{picture}
\end{subfig}

\hspace{2cm}

\begin{subfig}{b)}
\begin{picture}(150,46)(-3,-16)
\put(0,0){\chain{6}{$L_1$}{$b_1'$}{$b_1''$}}
\put(90,0){\chain{5}{$L_2$}{$b_2'$}{$b_2''$}}
\put(0,0){\line(0,1){30}}
\put(142,0){\line(0,1){30}}
\put(62,0){\line(1,0){28}}
\put(0,30){\line(1,0){142}}
\end{picture}
\end{subfig}

\hfill}

\caption{Reconnecting closed chains.}\label{fig_loops}
\end{figure}
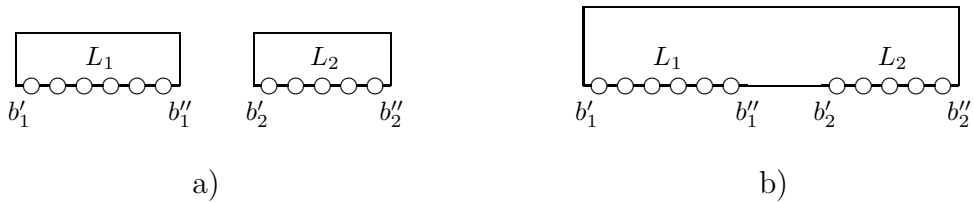

The following argument justifies eq.~(\ref{loop_condition}). An open chain of
length $L$ can be described by an effective Hamiltonian which only includes
boundary modes. If $\calM(H)=-1$, there are Majorana operators $b',b''$
associated with the ends of the chain. The parity operator $P$ (see
eq.~(\ref{parity_operator})) can be replaced by $s(L)(-ib'b'')$, where
$s(L)=\pm 1$. Thus the fermionic parity of $\ket{\psi_\alpha}$ is
$s(L)\,(-1)^\alpha$,\, $\alpha=0,1$. If we close the chain, the effective
Hamiltonian is $H_{\rm eff}(L)=\frac{i}{2}\,u\,b''b'$. (We have chosen to
write $b''b'$ in this order because $b''$ precedes $b'$ in the left-to-right
order on the loop, where they are next to each other). The parameter $u$
represents direct interaction between the chain ends (unlike $t$ from
eq.~(\ref{effective_Hamiltonian})), so $u$ does not depend on $L$. The ground
state of the closed chain is $\ket{\psi_1}$ if $u>0$ and $\ket{\psi_0}$ if
$u<0$.  Hence
\[
P(H(L))\,=\,-s(L)\,\sgn u.
\]

Now let us take two chains, one of length $L_1$, the other of length
$L_2$. There are two ways to close them up, see fig.~\ref{fig_loops}.
Both cases can be described by effective Hamiltonians:
\[
H_{\rm eff}(L_1)\oplus H_{\rm eff}(L_2)\,=\,
\frac{i}{2}\,u\,(b_1''b_1'+b_2''b_2'),
\qquad\
H_{\rm eff}(L_1+L_2)\,=\,
\frac{i}{2}\,u\,(b_1''b_2'+b_2''b_1').
\]
It follows that
\[
P(H(L_1))\,P(H(L_2)) \,=\, s(L_1)\,s(L_2), \qquad\quad
P(H(L_1+L_2)) \,=\, -\,s(L_1)\,s(L_2).
\]
So the equation~(\ref{loop_condition}) holds for $\calM=-1$. It also obviously
holds for $\calM=1$ because in this case there are no boundary modes to worry
about.

Computing the Majorana number in general (especially for strongly correlated
systems) may be a difficult task. However, the computation can be carried
through for any system of non-interacting electrons. Consider a periodic chain
of $L$ unit cells with $n$ fermionic sites (i.\,e.\ $2n$ Majorana operators)
per cell, which totals to $N=nL$ fermionic sites. We will index the Majorana
operators as $c_{l\alpha}$, where $l=1,\ldots,L$,\, $\alpha=1,\ldots,2n$. The
Hamiltonian is
\begin{equation}\label{periodic_Hamiltonian}
H \,=\, \frac{i}{4}
\sum_{l,m}\sum_{\alpha,\beta} B_{\alpha\beta}(m-l)\,c_{l\alpha}c_{m\beta}
\qquad\
\Bigl(B_{\alpha\beta}(j)^*=B_{\alpha\beta}(j)=-B_{\beta\alpha}(-j)\Bigr).
\end{equation}
We assume that the chain forms a loop, so $m-l$ should be taken$\pmod{L}$.

Eq.~(\ref{periodic_Hamiltonian}) is a special case
of~(\ref{quadratic_Hamiltonian}), so we will first find $P(H)$ for the general
quadratic Hamiltonian~(\ref{quadratic_Hamiltonian}), assuming that the matrix
$A$ is not degenerate. The canonical form of this
Hamiltonian~(\ref{canonical_form}) has an even ground state $\ket{0}$. The
transformation~(\ref{reduction_to_canonical}) can be represented as
conjugation by the parity-preserving unitary operator
$U=\exp\Bigl(\frac{1}{4}\sum_{l,m}D_{lm}c_lc_m\Bigr)$ if $W$ has the form
$W=\exp(D)$ for some real skew-symmetric matrix $D$, i.\,e.\ if $\det
W=1$. Otherwise, the transformation~(\ref{reduction_to_canonical}) changes the
parity. Hence
\begin{equation}\label{PH}
P(H)\,=\,\sgn\det W\,=\,\sgn\Pf A.
\end{equation}

We remind the reader that the Pfaffian $\Pf$ is a function of a
skew-symmetric matrix such that $(\Pf A)^2=\det A$. It is defined as follows
\begin{equation}\label{Pfaffian}
\Pf A \,=\, \frac{1}{2^{N}N!}\sum_{\tau\in S_{2N}}
\sgn(\tau)\,A_{\tau(1),\tau(2)}\cdots A_{\tau(2N-1),\tau(2N)}.
\end{equation}
(Here $S_{2N}$ is the set of permutations on $2N$ elements). For example,
\[
\Pf \left( \begin{array}{cccc}
0&a_{12}&a_{13}&a_{14}\\
-a_{12}&0&a_{23}&a_{24}\\
-a_{13}&-a_{23}&0&a_{34}\\
-a_{14}&-a_{24}&-a_{34}&0
\end{array}\right)\, =\, a_{12}a_{34}+a_{14}a_{23}-a_{13}a_{24}.
\]
In eq.~(\ref{PH}) we have used this property of the Pfaffian:
\begin{equation}\label{Pfaffian_property}
\Pf(WAW^T)\,=\,\Pf(A)\,\det(W).
\end{equation}

Now we are to compute the Pfaffian of the matrix $B$ from
eq.~(\ref{periodic_Hamiltonian}). First, we use the Fourier transform,
\begin{equation}\label{Fourier_transform}
{\tilde B}_{\alpha\beta}(q) \,=\,
\sum_{j} e^{iqj}B_{\alpha\beta}(j), \qquad\
q=2\pi\frac{k}{L}\pmod{2\pi},\,\ k=0,\ldots,N-1.
\end{equation}
The matrix ${\tilde B}(q)$ has these symmetries:
\begin{equation}
{\tilde B}^\dagger(q)=-{\tilde B}(q)={\tilde B}^T(-q).
\end{equation}
The spectrum $\epsilon(q)$ is a continuous real $2n$-valued function on a
circle (real numbers$\pmod{2\pi}$) given by the eigenvalues of $i{\tilde
B}(q)$. It has the symmetry $\epsilon(-q)=-\epsilon(q)$. The energy gap
assumption implies that $\epsilon(q)$ never passes $0$. It follows that there
are $n$ positive and $n$ negative eigenvalues for any $q$. Indeed, this is the
case for $q=0$ due to the $\epsilon\mapsto-\epsilon$ symmetry, hence it is
true for any $q$ by continuity.

It follows from eqs.~(\ref{Fourier_transform}) and~(\ref{Pfaffian_property})
that
\begin{equation}
\Pf B \,=\,
\left( \prod_{q=-q} \Pf {\tilde B}(q) \right)
\left( \prod_{q\not=-q} \det {\tilde B(q)} \right).
\end{equation}
Remember that $q$ is considered$\pmod{2\pi}$, so $q=-q$ when $q=0$ or
$q=\pi$. In the $q\not=-q$ case, each $\{q,-q\}$ pair is counted once.  Note
that $\det{\tilde B(q)}$ is a positive number since $i{\tilde B}(q)$ has $n$
positive and $n$ negative eigenvalues.  Hence
\begin{equation}
\sgn\Pf B \,=\, \prod_{q=-q} \sgn(\Pf {\tilde B}(q)) \,=
\left\{\begin{array}{ll}
\sgn(\Pf {\tilde B}(0))\, \sgn(\Pf {\tilde B}(\pi)) & \mbox{if $L$ is even,}
\smallskip\\
\sgn(\Pf {\tilde B}(0)) & \mbox{if $L$ is odd.}\\
\end{array}\right.
\end{equation}
Finally, we get
\begin{equation}\label{general_condition}
\calM(H) \,=\, \sgn(\Pf {\tilde B}(0))\, \sgn(\Pf {\tilde B}(\pi)).
\end{equation}

This very general equation can be simplified if superconductivity is a weak
effect, i.\,e.\ $|\Delta|\ll|\epsilon(0)|,\,|\epsilon(\pi)|$. Indeed, the
right hand side of~(\ref{general_condition}) makes perfect sense for a
$U(1)$-symmetric Hamiltonian
\begin{equation}
H_0 \,=\, \frac{1}{2}
\sum_{l,m}\sum_{\alpha,\beta} C_{\alpha\beta}(m-l)\,
a_{l\alpha}^{\dagger}a_{m\beta}
\qquad\quad
\Bigr(  C_{\alpha\beta}(j)^*=C_{\beta\alpha}(-j) \Bigr),
\end{equation}
where $\alpha,\beta=1,\ldots n$ refer to fermionic sites. The eigenvalues of
${\tilde C}(q)$ (the Fourier transform of $C$) form a ``single spectrum''
$\epsilon_0(q)$. The ``double spectrum'' defined above is
$\epsilon(q)=\pm\epsilon_0(q)$. It is easy to show that $\Pf {\tilde
B}(q)=\det {\tilde C}(q)$ for $q=0,\pi$. Hence
\begin{equation}\label{special_condition}
\calM(H_0)=(-1)^{\nu(\pi)-\nu(0)},
\end{equation}
where $\nu(q)$ is the number of negative eigenvalues of ${\tilde C}(q)$. Note
that $\nu(\pi)-\nu(0)$ equals$\pmod{2}$ the number of Fermi points on the
interval $[0,\pi]$. (A Fermi point is a point where $\epsilon_0(q)$ passes
$0$). In the most interesting case $\nu(\pi)-\nu(0)=1\pmod{2}$, the
Hamiltonian $H_0$ has a gapless spectrum. So eq.~(\ref{special_condition}) is
only relevant in the presence of superconductivity, i.\,e.\ a small
symmetry-breaking perturbation which opens an energy gap.

\section{Speculations about physical realization}
\label{sec_realization}

\begin{figure}[ht]
\centerline{\epsfbox{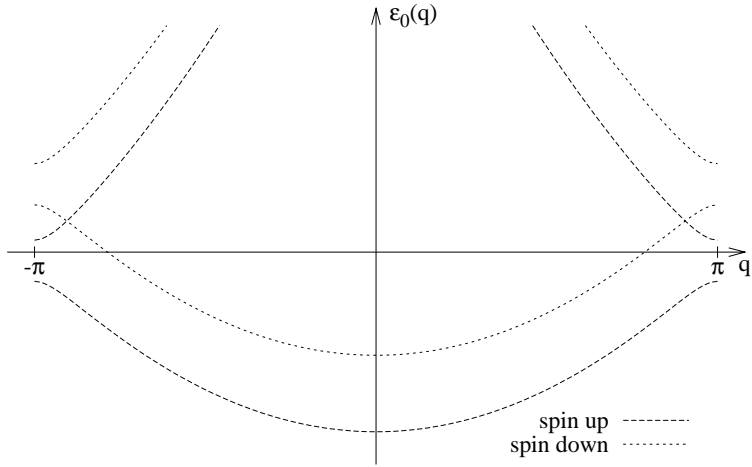}}
\caption{An electron spectrum in the presence of magnetic field and CDW.}
\label{fig_spectrum}
\end{figure}

Physical realization of an $\calM=-1$ quantum wire is a difficult task because
electron spectra are usually degenerate with respect to spin, so $\nu(0)$ and
$\nu(\pi)$ are even. The degeneracy at $q=0$ and $q=\pi$ can be lifted only if
the time reversal symmetry is broken. Thus spin-orbit interaction does not
help. External magnetic field could help, but the Zeeman energy $g\mu_B{\cal
H}$ is usually small compared to other spectrum parameters, so $\nu(0)$ and
$\nu(\pi)$ do not change. The situation may be different for charge and spin
density waves which add fine features to the electron spectrum.  Charge
density waves (CDW) tend to occur at the wave vector $q_*=2q_F$ so that a gap
opens at the Fermi level. In the presence of magnetic field, $q_F$ is slightly
different for the $\uparrow$ and $\downarrow$ spin components, so it is
possible that $q_*$ matches only one of them. The resulting spectrum is shown
in fig.~\ref{fig_spectrum} in the $q_*/(2\pi)$ units. This scenario can be
realized if $|\Delta|\lesssim E_{\rm CDW}\lesssim g\mu_B{\cal H}$.

Another speculative possibility is to use midgap states at the edge of a
two-dimensional $p$-wave superconductor~\cite{midgap}.

\begin{figure}[ht]

\sbox{\TempBox}{%
\begin{picture}(0,0)
\put(0,0){\line(1,0){40}}
\put(0,10){\line(1,0){40}}
\put(10,30){\line(1,0){40}}
\put(0,0){\line(0,1){10}}
\put(40,0){\line(0,1){10}}
\put(50,20){\line(0,1){10}}
\put(0,10){\line(1,2){10}}
\put(40,10){\line(1,2){10}}
\put(40,0){\line(1,2){10}}
\end{picture}}

\newcommand{\brick}[1]{%
\TempLength=40\unitlength%
\begin{picture}(0,0)
\put(0,0){\usebox{\TempBox}}
\put(0,2){\hbox to \TempLength {\footnotesize\hfil#1\hfil}}
\end{picture}}

\hbox to \textwidth{\hfill

\begin{subfig}{a)}
\begin{picture}(97,30)
\put(0,0){\brick{$\theta_1$}}
\put(47,0){\brick{$\theta_2$}}
\put(8,19.7){\thicklines\line(1,0){81}}
\put(8,20.3){\thicklines\line(1,0){81}}
\end{picture}
\\[7mm]
$\theta=\theta_2-\theta_1$
\vspace{4mm}
\end{subfig}

\hfill

\begin{subfig}{b)}
\epsfbox{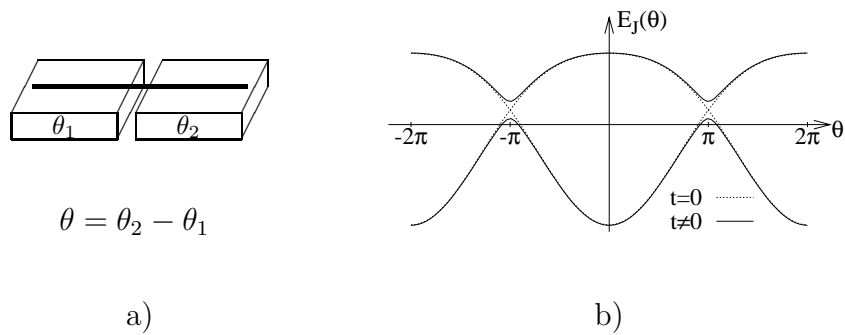}
\end{subfig}

\hfill}

\caption{A Josephson junction made of quantum wire.}
\label{fig_junction}
\end{figure}

A quantum wire bridge between two superconducting leads (see
fig.~\ref{fig_junction}a) could be used as an experimental test for Majorana
fermions. When the phase parameter $\theta_2$ in the right piece of
superconductor changes by $2\pi$ (relative to $\theta_1$), a fermionic
quasiparticle is effectively transported to the junction region. At the same
time, the Majorana fermions at the ends of the wire switch from $\ket{\psi_0}$
to $\ket{\psi_1}$ or vice versa. If the quasiparticle stays localized, the
junction parameters change. They change back when $\theta_2$ changes by
another $2\pi$. Thus the Josephson current is $4\pi$-periodic as a function of
$\theta=\theta_2-\theta_1$. In fact, it is more accurate to say that the
Josephson energy $E_J$ is $2\pi$-periodic but $2$-valued, as shown in
fig.~\ref{fig_junction}b. The two levels may not quite cross at $\theta=\pi$
due to a non-vanishing tunneling amplitude $t\propto e^{-L/l_0}$, where $L$ is
the distance between the junction and the closest end of the wire.

Interesting phenomena can also take place in the simple layout shown in
fig.~\ref{fig_layout}. Suppose that the superconducting island supporting the
quantum wire is connected to a larger piece of superconductor through an
ordinary Josephson junction. If the Coulomb energy is comparable to the
Josephson energy, spontaneous phase slips can occur. Each $2\pi$ phase slip is
accompanied by the operator $V$ (see eq.~\ref{V}). The phase slips occur by
tunneling, so the effective Hamiltonian is
\begin{equation}
H_{{\rm eff.}1}=-\lambda V- \lambda^* V^\dagger =
\frac{i}{2}s(L)\,t\,b'b'',\qquad\quad
t=4\mathop{\rm Re}\lambda,
\end{equation}
where $\lambda$ is the amplitude of the $\theta\mapsto\theta+2\pi$ process
while $\lambda^*$ corresponds to the reverse process. Similarly, if the
superconducting island supports two quantum wires, the effective Hamiltonian
becomes
\begin{equation}
H_{{\rm eff.}2}=-\lambda V_1V_2- \lambda^* V_1^\dagger V_2^\dagger =
\frac{1}{2}s(L_1)s(L_2)\,t\,b_1'b_1''b_2'b_2''.
\end{equation}
Turning $\lambda$ on and off can be possibly used for quantum gates
implementation.

\bigskip\noindent {\bf Acknowledgements.}  I am grateful to J.\,Preskill,
M.\,Feigelman, P.\,Vigman and V.\,Yakovenko for interesting discussions.

\end{document}